# $\mathcal{PT}$-symmetry in honeycomb photonic lattices

**Alexander Szameit, Mikael C. Rechtsman, Omri Bahat-Treidel, and Mordechai Segev**

*Physics Department and Solid State Institute, Technion, 32000 Haifa, Israel*

**Abstract**

We apply gain/loss to honeycomb photonic lattices and show that the dispersion relation is identical to tachyons – particles with imaginary mass that travel faster than the speed of light. This is accompanied by $\mathcal{PT}$-symmetry breaking in this structure. We further show that the $\mathcal{PT}$-symmetry can be restored by deforming the lattice.

Honeycomb photonic lattices [1] - the optical version of the well known graphene [2] (a monolayer of carbon atoms arranged in a honeycomb geometry) - shares many common features with that unique material. Interestingly, such "photonic graphene" actually displays several additional phenomena that are not manifest in the original electronic system, for example, nonlinear conical diffraction [3]. When transferring the physics of particles onto an analogous optical setting, which is justified by the fundamental wave-particle duality, the observer benefits from spatial (rather than temporal) evolution and from almost arbitrary scalability of the length scale of the refractive index contrast, compared to the atomic potential. Furthermore, the wave function can be directly imaged and monitored as it evolves. In this vein, it was recently suggested to study Klein tunnelling in honeycomb photonic lattices [4]. Importantly, optical systems offer the possibility of employing nonlinearity, which has significant impact on the wave packet evolution [3, 5].

In a different area, complex non-conservative systems, which, under special conditions, may exhibit non-Hermitian but $\mathcal{PT}$-symmetric Hamiltonians, have garnered growing scientific interest in recent years [6]. $\mathcal{PT}$-symmetric systems are characterized by a complex potential, which has neither parity symmetry ($\mathcal{P}$) nor time-reversal symmetry ($\mathcal{T}$), yet the Hamiltonian commutes with $\mathcal{PT}$, and both share all of their eigenstates. Under these conditions, the eigenvalues of the Hamiltonian are real, in spite of the fact that the potential is complex [6]. This concept has motivated an ongoing debate in several forefronts of physics on the impact of $\mathcal{PT}$-symmetry, such as quantum field theories [7], non-Hermitian Anderson models [8] and open quantum systems [9]. However, the concept of $\mathcal{PT}$-symmetry was also introduced into the domain of optics, which provides exceptional conditions to observe $\mathcal{PT}$-symmetric systems [10]. It turned out that $\mathcal{PT}$- symmetry plays an important role in the light evolution in optical systems [11]. Their simplest realization occurs for two identical coupled waveguides, one with gain and the other with loss, such that the real part of the refractive index is symmetric with respect to the interchange of waveguides whereas the imaginary counterpart is anti-symmetric. This realization was recently demonstrated in experiments [12], although some of the underlying effects were observed even without gain, when both waveguides contain unequal loss [13].

Here, we show that adding spatially-alternating gain/loss to a photonic honeycomb lattice – small as it may be – breaks parity and the $\mathcal{PT}$-symmetry. However, the system gives rise to a dispersion relation that resembles that of tachyons –hypothetical particles with pure imaginary mass and a group velocity exceeding the vacuum speed of light [14]. Nevertheless, applying appropriate strain to the honeycomb lattice restores $\mathcal{PT}$-symmetry.

The structure of complex honeycomb waveguides lattices with alternating gain/loss is sketched in Fig. 1a. In our lattice we allow a deformation, yielding non-equal coupling constants $c_1 = ct$, $c_2 = c_3 = c$, with $t=1$ as the unperturbed structure [4]. Additionally, one may introduce a detuning $\Delta$ in the effective index between adjacent waveguides. A honeycomb lattice is composed of two displaced hexagonal sublattices $a$ and $b$. Therefore, in a tight-binding model, the dynamics of the entire system can be described by the equations [3-5]

$$i\partial_z a_{m,n} = -\Delta a_{m,n} - i\gamma a_{m,n} + c(tb_{m-1,n} + b_{m,n+1} + b_{m,n-1})$$
$$i\partial_z b_{m,n} = +\Delta b_{m,n} + i\gamma b_{m,n} + c(ta_{m+1,n} + a_{m,n+1} + a_{m,n-1})$$
(1)

where $c$ is the coupling constant between adjacent guides. The quantity $\gamma$ describes the gain and loss of the waveguides in the $a$ and $b$ sublattices. The spectrum of the eigenvalues $\beta(\mu, \nu)$ can be obtained by substituting the plane wave solutions $a_{m,n} = A\exp\{i(\beta z + \sqrt{3}\mu m + \nu n)\}$, and $b_{m,n} = B\exp\{i(\beta z + \sqrt{3}\mu m + \nu n)\}$ into Eq. (1), where $\mu$ and $\nu$ are dimensionless in units of the inverse inter-site spacing and $(\mu,\nu)$ represents the transverse wave vector, $A$ and $B$ are amplitudes, and the factor $\sqrt{3}$ results from the lattice structure. This results in the eigenvalue problem

$$\begin{pmatrix} \Delta + i\gamma & cte^{-i\sqrt{3}\mu} + 2c\cos\nu \\ cte^{i\sqrt{3}\mu} + 2c\cos\nu & -\Delta - i\gamma \end{pmatrix}\begin{pmatrix} A \\ B \end{pmatrix} = \beta\begin{pmatrix} A \\ B \end{pmatrix}, \qquad (2)$$

yielding the dispersion relation for the propagation constant $\beta$, which describes the rate of phase evolution in the propagation direction:

$$\beta = \pm\sqrt{\Delta^2 - \gamma^2 + 2i\gamma\Delta + c^2t^2 + 4c^2\cos^2\nu + 4tc^2\cos\nu\cos\sqrt{3}\mu}. \qquad (3)$$

An ideal (unperturbed) honeycomb lattice is obtained for $\Delta=\gamma=0$ and $t=1$, and its dispersion relation is shown in Fig. 1b. One of the striking features of the band structure of this system is the existence of the so-called Dirac region in the vicinity of the intersection points (the vertices) between the first and the second bands. In this regime the Hamiltonian of the system – defined in Eq. (2) – can be expanded into a Taylor series [15], resulting in a mathematical structure similar to the one in the relativistic Dirac equation, which describes relativistic quantum particles:

$$H = f(t)\tilde{\nu}\sigma_1 + ct\sqrt{3}\tilde{\mu}\sigma_2 + (\Delta + i\gamma)\sigma_3. \qquad (4)$$

Here, $\sigma_{1,2,3}$ are the Pauli matrices:

$$\sigma_1 = \begin{pmatrix} 0 & 1 \\ 1 & 0 \end{pmatrix};\quad \sigma_2 = \begin{pmatrix} 0 & -i \\ i & 0 \end{pmatrix};\quad \sigma_3 = \begin{pmatrix} 1 & 0 \\ 0 & -1 \end{pmatrix} \qquad (5)$$

and $f(t) = 2c\sqrt{1 - t^2/4}$. The quantities $\tilde{\mu} = \mu - \mu_0$ and $\tilde{\nu} = \nu - \nu_0$ are the components of the transverse wave vector $(\tilde{\mu},\tilde{\nu})$ measured from the position of the given Dirac point $[\mu_0,\nu_0]$. Note that the regime of validity of this expansion is $t<2$ [3]. In Eq. (4), the detuning $\Delta$ plays the role of a mass of a relativistic fermion in Dirac's theory, whereas the gain/loss factor $\gamma$ represents an *imaginary* mass. The Dirac region for $\Delta=\gamma=0$, resembling the dispersion relation of a massless particle, is shown in Fig. 1(c). In contrast, when both sublattices *a* and *b* are detuned by $\Delta>0$, a gap opens between the two bands, as shown in Fig. 1(d), where $\Delta=0.2$. The magnified Dirac region (see Fig. 1(e)) reveals a dispersion relation in the form of a double sheeted hyperboloid, indicating the resemblance of the dispersion of a massive relativistic particle.

The situation changes drastically when gain and loss are introduced. Whereas for $\gamma=0$ the Hamiltonian is Hermitian, resulting in a purely real eigenvalue spectrum, a gain/loss structure with $\gamma>0$ results in a non-Hermitian Hamiltonian that in general exhibits a non-real eigenvalue spectrum. Exactly this happens in a honeycomb photonic lattice, where according to Eq. (3) one finds that (if $\Delta=0$ and $t=1$) for every $\gamma$ there exist imaginary $\beta$ in the Dirac region. In more technical terms, complex eigenvalues of the Hamiltonian appear when $\mathcal{PT}$ and the Hamiltonian do not share all of their eigenvectors. Such a system is said to have broken $\mathcal{PT}$-symmetry, although the $\mathcal{PT}$ operator still commutes with the Hamiltonian. This seeming paradox stems from the fact that the $\mathcal{T}$ operator is an anti-linear operator. A graph of the real part of the dispersion relation with $\gamma=0.5$ is shown in Fig. 2a, and a magnified plot of the Dirac region is shown in Fig. 2b. In contrast to "conventional" graphene, the real part of the dispersion relation is now a single sheeted hyperboloid. Figures 2c,d show the imaginary part of the dispersion relation and a zoom-in of the Dirac region, indicating purely imaginary eigenvalues around the original vertices. Note that there is no detuning between the sublattices *a* and *b*, i.e. $\Delta=0$. If the structure exhibits both detuning and gain/loss (i.e. $\Delta>0$, $\gamma>0$) then, according to Eq. (3) all eigenvalues are complex due to the term $2i\gamma\Delta$.

Interestingly, Eq. (4) suggests that our system resembles the dynamics of relativistic particles with imaginary rest mass – generally known as tachyons [14]. Tachyons are hypothetical particles that exhibit various peculiar features; the most striking is that they travel faster than the vacuum speed of light. Furthermore, these strange particles get faster the lower their energy is, and approach an infinite velocity when their energy is zero. Note that – had tachyons been conventional, localizable, particles – they could be used to send signals faster than light, which would, however, lead to violations of causality. A full discussion of applications for tachyons can be found in [16]. Currently, according to the contemporary and widely accepted understanding of the concept of a particle, tachyon particles are assumed to be non-existent [17]. Nevertheless, despite the theoretical arguments against the existence of tachyons, no clear experimental evidence for or against their existence has been found [18].

A honeycomb photonic lattice with a gain/loss structure provides a classical optical analogue of tachyons, which can be experimentally probed in a table-top experiment. First evidence is given by the fact that the gain/loss factor $\gamma$ can be interpreted as an imaginary mass due to its appearance on the main diagonal of the Dirac Hamiltonian in Eq. (4). Additionally, an intuitive understanding of the analogous behavior of an optical wave packet in photonic graphene and tachyons in vacuum can be gathered by analyzing the hyperbolic (single sheeted) dispersion relation. In Fig. 3a, we show a cross section of the real part of the dispersion relation through the Dirac region. The transverse velocity of a wave packet, which is defined as the gradient of the dispersion relation and is equivalent to the group velocity [19], increases strongly above the values for "conventional" photonic graphene in the Dirac region, and diverges when the propagation constant $\beta$ approaches zero (see Fig. 3b). Note that the propagation constant $\beta$ of an optical wave packet is the optical analogue of the quantum mechanical energy of an evolving particle. Therefore, a wave packet (beam) associated with the regions where $\beta \rightarrow 0$ will travel with a transverse group velocity far beyond the speed of corresponding wave packets in "conventional" honeycomb photonic lattices.

We test our analytical results in numerical simulations by evaluating the displacement of a wave packet after some propagation, as a function of the transverse wave number $\mu$. The results are summarized in Fig. 3c, comparing the final position of a propagated wave packets in the cases $\gamma=0$ and $\gamma=0.5$. Indeed, in the gain/loss system the transverse displacement of the wave packet is larger. In Fig. 3d, the transverse speed of a propagating wave packet is plotted as a function of the initial transverse momentum, showing the predicted tachyonic behavior: in the Dirac region, the transverse speed of the wave packet in the gain/loss system clearly exceeds the transverse velocity of a wave packet in conventional honeycomb photonic lattice, defined by the constant slope of the linear dispersion curve. The wave packet travels faster than the optical analogue of the speed of light in the Dirac equation. Note that close to the point where $\beta$ is close to zero and becomes imaginary, the numerically obtained group velocity of the wave packet deviates from the analytic prediction. We attribute this fact to the finite width of our wave packet: the velocities diverge in a single point in reciprocal space only, in a region of extremely low density of states. Our wave packet is very broad but is still finite, hence its spectrum spans over some region, which, small as it may be, is not a single point. Consequently, the wave packet used in the simulations inevitably displays some average over this region in momentum space. Nevertheless, our system clearly exhibits beams propagating at angles steeper than dictated by the underlying honeycomb lattice – i.e. the group velocity exceeds the upper bounds in conventional honeycomb lattice – which can be studied in table-top experiments.

Finally, we point out that the strain $t$, as defined in Eq. (1), is capable of restoring the $\mathcal{PT}$-symmetry in complex honeycomb photonic lattices. In conventional lattices ($\Delta=\gamma=0$), increasing the strain forces the Dirac points to move towards each other until they merge at $t=2$ [15, 20]. Using Eq. (3), one can easily show that, for vanishing detuning ($\Delta=0$), for any given gain/loss factor $\gamma$ - all eigenvalues of $\beta$ become real above a threshold strain $t \geq 2+(\gamma/c)$. Therefore, in such a setting, the structure is fully $\mathcal{PT}$-symmetric, and critical points appear exactly for $t=2+(\gamma/c)$. The Hamiltonian of this system reads [3]

$$H = \left[2(t-2) - 3ct\tilde{\mu}^2 + c\tilde{v}^2\right]\sigma_1 + ct\sqrt{3}\tilde{\mu}\sigma_2 + (\Delta + i\gamma)\sigma_3. \tag{6}$$

Note that the quadratic term $\sim \tilde{\mu}^2$ may not be neglected since the dispersion obtained from Eq. (6) must coincide with the expansion of Eq. (3). In Fig. 4a, we show the dispersion relation for photonic graphene, with no detuning ($\Delta=0$), but with gain/loss structure ($\gamma=0.5$), and additional strain $t = 2+(0.5/c)$ compensating for the imaginary part of Eq. (3). In the magnified view of the Dirac region (see Fig. 4b), one finds that this particular strain yields a vertex at the center of the Dirac region and an anisotropic dispersion relation that is linear in $\mu$ and parabolic in $v$. However, for a slightly larger strain, namely $t = 2+(0.55/c)$ - all eigenvalues $\beta$ remain real, but a gap opens between both bands, as depicted in Fig. 4c. This is particularly visible in a magnified view on the Dirac region, shown in Fig. 4d. Hence, a beam with eigenvalues close to the vertex will propagate with constant amplitude in the $\mathcal{PT}$-symmetric system, in spite of the fact that gain and loss are present in the system, whereas in the unstrained system the beam will experience an amplitude growth beyond all limits due to the imaginary eigenvalues, as shown in Fig. 5.

In conclusion, we have described the properties of complex honeycomb photonic lattices and revealed its potential to support "optical tachyons". Whereas $\mathcal{PT}$-symmetry is always broken in undeformed complex honeycomb lattices, all eigenvalues may be rendered real by introducing a lattice deformation of the proper value, thereby restoring the $\mathcal{PT}$-symmetry. Besides the potential of using honeycomb photonic lattices as a classical simulator for relativistic physics [21] complimentary to quantum simulators [22], we envision also various applications, in particular for dielectric nonreciprocal optical elements, with no net loss.

This work was sponsored by an Advanced Grant from the European Research Council (ERC), by the Israel Science Foundation, and by the German-Israeli Foundation (GIF). AS acknowledges support by the German Academy of Science Leopoldina (grant LPDS 2009-13). MCR is grateful to the Azrieli foundation for the award of an Azrieli fellowship.

# Figure Captions:

**Fig. 1.**

(a) Sketch of complex photonic graphene. The red and blue waveguides exhibit gain and loss, respectively. Additionally, a strain can be applied along the horizontal direction. (b) The dispersion relation of conventional photonic graphene, where $\Delta=\gamma=0$. (c) The magnified conical Dirac region, resembling the dispersion relation for a massless relativistic particle. (d) The dispersion relation of photonic graphene, where $\gamma=0$, but $\Delta=0.2$. (e) The magnified Dirac region, resembling the dispersion relation for a massive relativistic particle.

**Fig. 2:**

(a) The real part of the dispersion relation of photonic graphene for $\gamma=0.5$ and $\Delta=0$. (b) A magnified view on the hyperbolic shape of the real part of $\beta$, which is described by a Dirac equation for particles with imaginary mass. (c) A plot of the imaginary part of the dispersion relation, showing the eigenvalues $\beta$ become imaginary in the vicinity of the Dirac points. (d) A magnified view of the imaginary eigenvalues in the Dirac region.

**Fig. 3:**

(a) A cross section through the real part of the dispersion relation in Fig. 1(b) along $\mu$ for $\nu=0$. (b) The slope of (a), that diverges in the vicinity of the Dirac region, indicating transverse speeds exceeding that in conventional graphene. (c) Comparison of the initial position of the wave packet with momentum [$\mu=0$; $\nu=0.5\pi$] (black solid line) and the final position when $\gamma=0$ (red dotted line) and $\gamma=0.5$ (blue dashed line). (d) Numerical simulation of the transverse displacement of a wave packet as a function of the initial transverse momentum (red solid line) compared to the analytic solution (blue dashed line). Speeds exceeding the maximal value in conventional photonic graphene (black dotted line) correspond to tachyonic behavior.

**Fig. 4:**

(a) Dispersion relation for $\gamma=0.5$ and $t = 2 + (0.5/c)$. All eigenvalues $\beta$ are real, despite the complex potential, indicating that the structure exhibits $\mathcal{PT}$-symmetry. (b) The enlarged Dirac region, where a vertex exists and the dispersion is linear in both $\mu$ and $\nu$, but anisotropic. (c) Dispersion relation, when the strain is increased to $t = 2 + (0.55/c)$. All eigenvalues $\beta$ remain real, but a gap opens between both bands. (d) The enlarged Dirac region shows that, sufficiently away from the former vertex, the dispersion remains linear and anisotropic in both $\mu$ and $\nu$.

**Fig. 5:**

Comparison of the amplitude of a propagating wave packet for $\gamma=0.5$ with momentum [$\mu=0$; $\nu=0.6\pi$] and no strain $t=0$ (black solid line), and in the same region for $t=t_c$ when $\mathcal{PT}$-symmetry is restored and all eigenvalues are real (red dashed line).

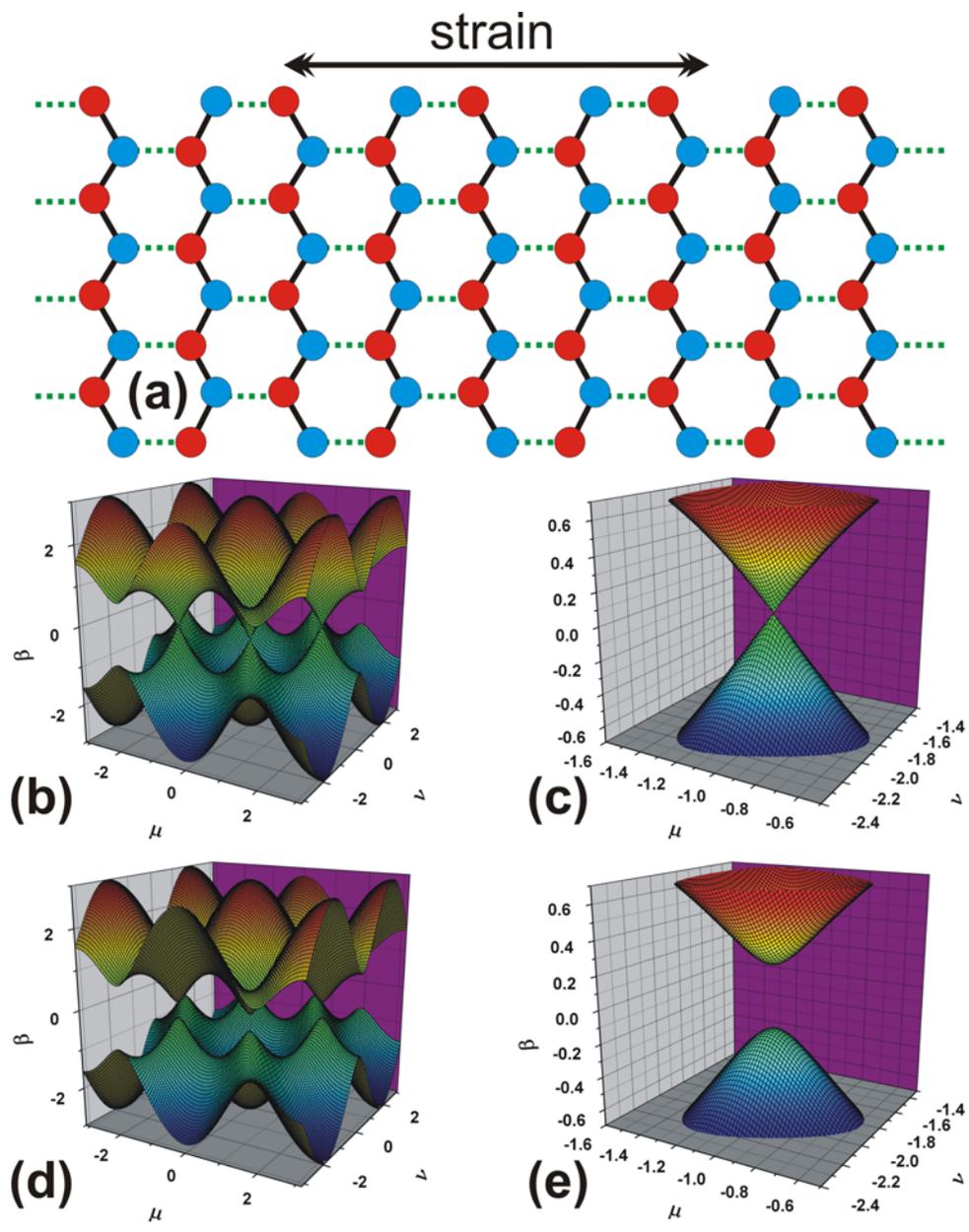

**Figure 1**

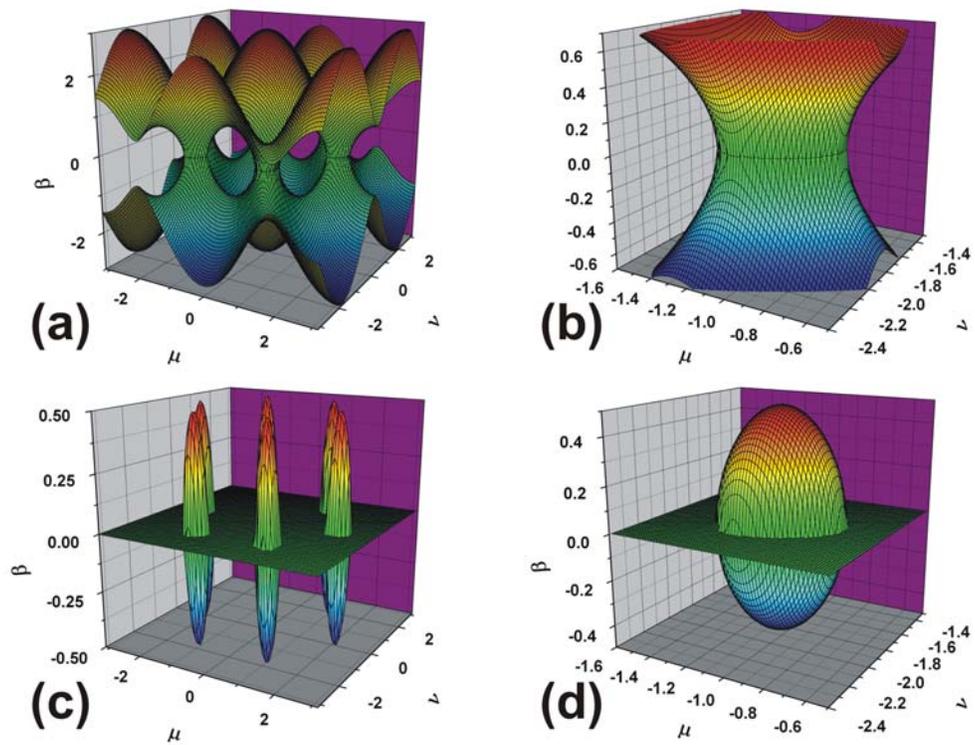

**Figure 2**

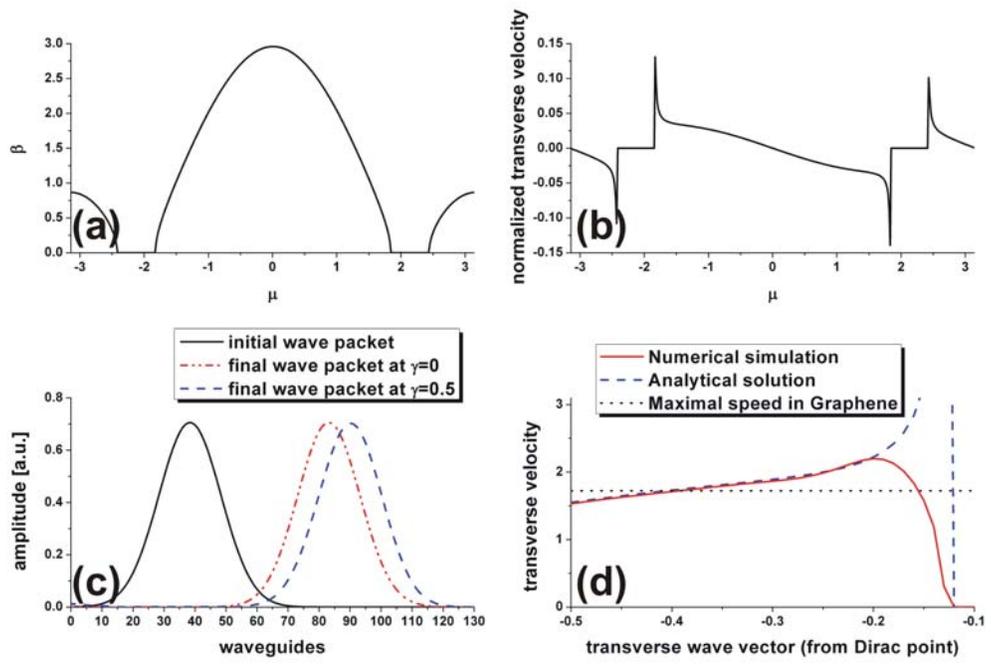

**Figure 3**

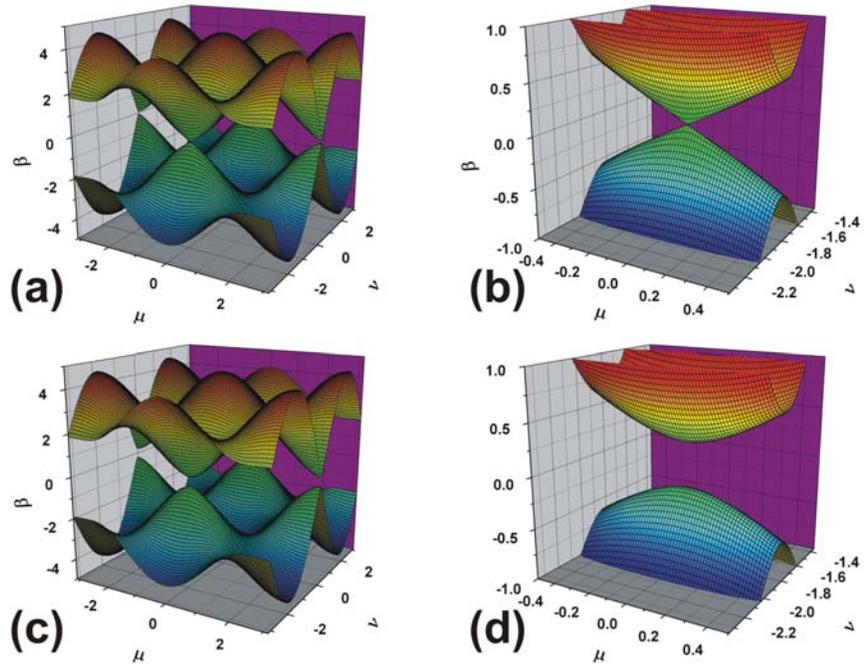

**Figure 4**

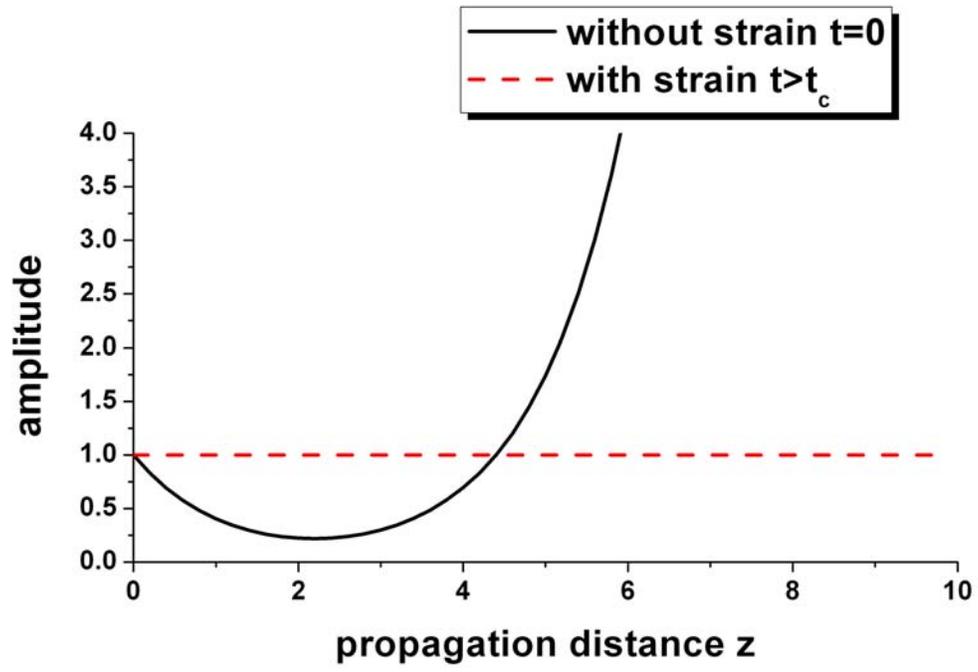

**Figure 5**